\documentclass{article}
\usepackage{amsmath,graphicx,mathtools,amssymb,setspace,lipsum,cuted,multicol,multirow,makecell,cite,spconf}
\usepackage{booktabs}
\usepackage{enumitem}
\usepackage{hyperref}
\usepackage{xurl}
\usepackage{url}
\usepackage{xcolor}

\let\OLDthebibliography\thebibliography
\renewcommand\thebibliography[1]{
  \OLDthebibliography{#1}
  \setlength{\parskip}{0pt}
  \setlength{\itemsep}{0pt plus 0.3ex}
}

\title{KaraSinger: Score-Free Singing Voice Synthesis with VQ-VAE using Mel-spectrograms}
\name{Chien-Feng Liao$^{\ast}$\thanks{$\ast$  Equal contribution. JY Liu has moved to Earth Species Project.}, Jen-Yu Liu$^{\ast}$, and Yi-Hsuan Yang}
\address{
  Taiwan AI Labs, Taiwan \\
  \small \tt \{chienfeng.liao,~jyliu,~yhyang\}@ailabs.tw
}

\begin{document}
\ninept
\maketitle
\begin{abstract}
In this paper, we propose a novel neural network model called KaraSinger for a less-studied singing voice synthesis (SVS) task named \emph{score-free SVS}, in which the prosody and melody are spontaneously decided by machine. KaraSinger comprises a vector-quantized variational autoencoder (VQ-VAE) that compresses the Mel-spectrograms of singing audio to sequences of discrete codes, and a language model (LM) that learns to predict the discrete codes given the corresponding lyrics. For the VQ-VAE part, we employ a Connectionist Temporal Classification (CTC) loss to encourage the discrete codes to carry phoneme-related information. For the LM part, we use location-sensitive attention for learning a robust alignment between the input phoneme sequence and the output discrete code.
We keep the architecture of both the VQ-VAE and LM light-weight for fast training and inference speed. 
We validate the effectiveness of the proposed design choices using a proprietary collection of 550 English pop songs sung by multiple amateur singers. The result of a listening test shows that KaraSinger achieves high scores in intelligibility, musicality, and the overall quality.

\end{abstract}
\section{Introduction}
\label{sec:intro}

Singing voice synthesis (SVS) is the task of computationally generating singing voices from music scores and lyrics \cite{cook1996singing,cho2021survey}. Such a model aims at producing realistic and expressive vocals in which the contents are determined by the input lyrics and score, which specifies the pitch and duration for every melody note. According to the main approach taken, existing methods can be categorized into unit-selection based methods \cite{kenmochi2007vocaloid, bonada16_interspeech}, hidden Markov model (HMM)-based statistical parametric models \cite{saino2006hmm, oura2010recent}, or deep neural network-based SVS \cite{hono2018recent, kim2018korean, nakamura2020fast, ren2020deepsinger, chen2020hifisinger, lu20c_interspeech}. 
In light of the shared underlying pipelines, newly developed techniques for neural text-to-speech (TTS) \cite{tan2021survey} have also been incorporated to SVS. 
For example, sequence-to-sequence (seq2seq) models with content/location-based attentions, such as the Tacotrons \cite{wang17n_interspeech, 8461368}, have been adapted to SVS  \cite{lee2019adversarially, angelini20_interspeech, 9053944, gu2021bytesing}.

While existing SVS models can sing along nicely to a pre-assigned musical score, there is a barrier for end users to employ such models to create singing voices---while ordinary users can easily provide a sentence of lyrics as input to the model, they may have a hard time composing a reasonable melody as the other mandatory model input. Here, we propose to unleash the creativity of computers and explore the realm of \emph{score-free SVS} (or, text-to-music), 
where a model learns the prosody and melody of music implicitly from data at training time. At inference time, the model sings without the guidance of human input other than  lyrics. To our best knowledge, such a score-free SVS task has not been much studied.\footnote{Our previous work \cite{liu2019score, liu2020unconditional} proposed a score- and lyrics-free singing generation models using generative adversarial network (GAN). These models only need vocals for training, taking neither the score nor lyrics as input. What we study in this paper is a relatively more constrained setting, where the singing voice generating model is conditioned on lyrics but not on scores.}

The Jukebox model \cite{dhariwal2020jukebox} proposed by OpenAI stands out as the work most closely related to ours in the literature. This model can generate arbitrary musical audio (not just solo singing voices)  conditioning on an input lyrics. Specifically, it aims at producing long-range coherent raw waveform by compressing each waveform in the training data to discrete vector-quantized (VQ) codes using a multi-scale VQ-variational autoencoder (VAE) \cite{oord2017neural,kaiser18a,razavi2019generating,polyak21interspeech,walker21arxiv}, and then using an autoregressive Transformer \cite{vaswani2017attention} to model the sequence of VQ codes. During inference, the Transformer acts as a language model (LM) that predicts discrete VQ codes given genre, artist, and/or lyric as conditions. Finally, the VQ-VAE decoder converts the sequence of VQ codes back to the waveform domain. With massive amount of crawled songs (1.2 million) and enormous model complexity (over 200 Transformer layers), Jukebox can generate multi-instrument music with high fidelity over various musical 
styles. 

The primary contribution of this paper is the development of a lite Jukebox-like model tailored for generating singing-only audio, specifically a score-free seq2seq SVS model that capitalizes a Transformer encoder to take lyrics input and a  Transformer decoder to generate discrete codes of singing voices. While the original Jukebox model achieves impressive results, it suffers from word-skipping problems commonly seen in early TTS and SVS models, because the model was not designed for generating singing vocal.\footnote{This can be easily seen by choosing "Unseen lyrics" in their demo website, \url{https://jukebox.openai.com}.} Moreover, the original model is computationally prohibitive.

To address these issues, we propose the following modifications.
First, we utilize gated recurrent units (GRUs) and location-based attention \cite{chorowski2015attention} as our seq2seq architecture, which has been shown effective for TTS and SVS \cite{8461368, angelini20_interspeech}. 
Second, we find it crucial to add a Connectionist Temporal Classification (CTC) \cite{graves2006connectionist} loss to the VQ-VAE training process; without it the seq2seq model struggles to learn meaningful alignments in cross-attention to attend to the input lyrics. Finally, our model works on the Mel-spectrograms instead of raw waveforms. Due to the light-weight architecture, the proposed model takes only 4 days to train on a single NVIDIA RTX 2080Ti, and the inference time needed is approximately 1:1 to the length of the generated utterance.\footnote{In comparison, Jukebox is reported to take 2 weeks to train on hundreds of V100 GPUs, and 8 hours to synthesize a one-minute song \cite{dhariwal2020jukebox}.} 
We refer to our model as KaraSinger.\footnote{To our knowledge, DiscreTalk \cite{hayashi2020discretalk} is to date the only work that employs a Jukebox-like model for TTS. Our model differs from DiscreTalk in the task addressed (TTS vs score-free SVS), data representation (waveforms vs Mel-spectrograms), model architectures, and the objective functions.}


\begin{figure}[t]
\includegraphics[width=0.95\linewidth,keepaspectratio]{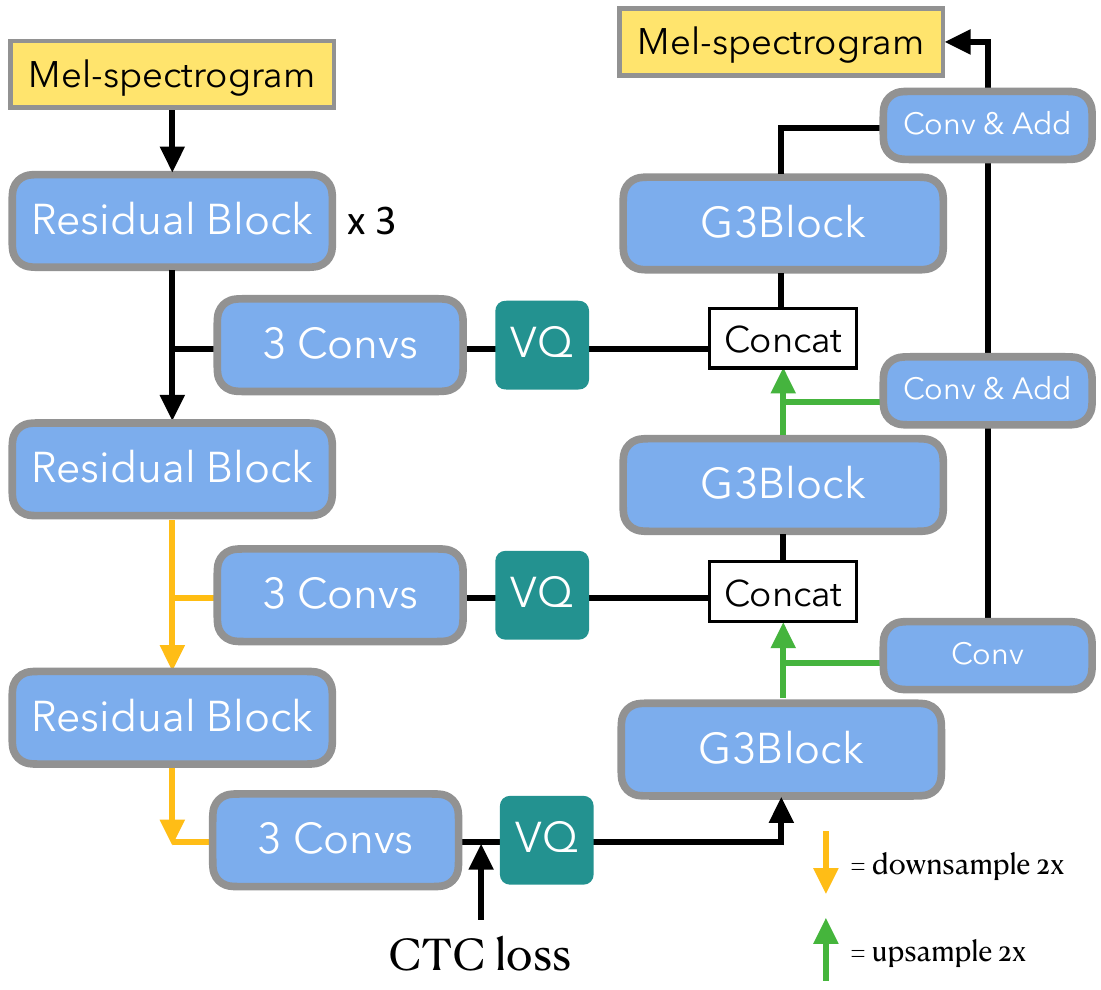}
\centering
\caption{Schematic diagram of the VQ-VAE part of our model. The `G3 Block' is adopted from our previous work \cite{liu2019score}, which comprises 1-D convolutions, group-normalization, and bi-directional GRUs.}
\label{fig:vqvae}
\end{figure}

\section{The Proposed Model}
\label{sec:system}
Instead of modelling the distribution of singing voices frame-by-frame in a continuous space, discrete latent models such as VQ-VAE \cite{oord2017neural,kaiser18a,razavi2019generating,polyak21interspeech,walker21arxiv} 
uses a discrete representations of data for generative modeling. Once the discrete space is constructed, the singing voices will be described as a sequence of ``tokens,'' like symbolic audio words. The LM underlying the quantized audio words can then be modeled by powerful autoregressive models such as Transformers. 

\begin{figure*}[t]
\includegraphics[width=.94\linewidth,keepaspectratio]{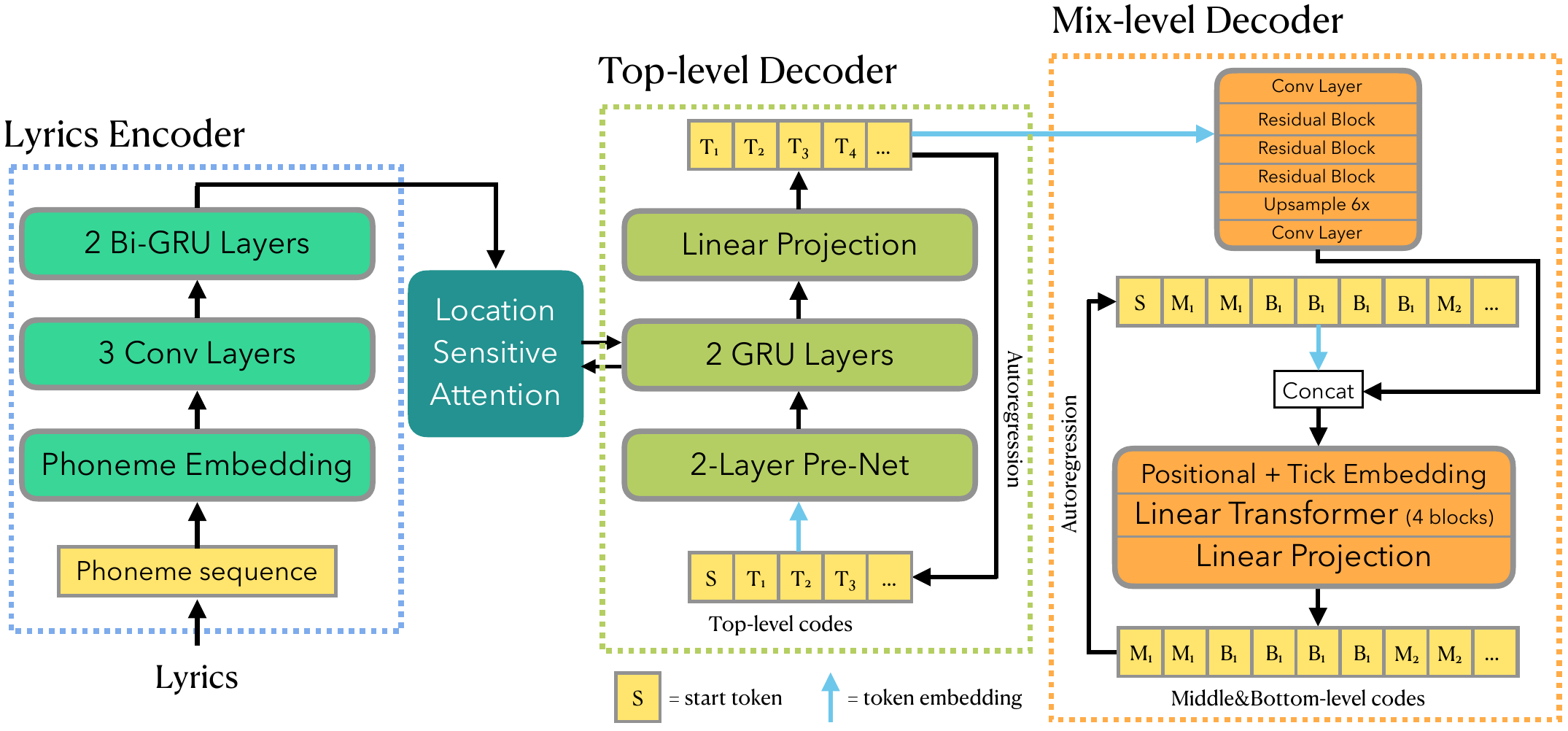}
\centering
\caption{The LM part of our model that models the three layers of VQ codes (\texttt{T}: top, \texttt{M}: middle, \texttt{B}: bottom). The combination of the lyrics encoder and the top-level decoder follows the same design as the acoustic model in Tacotron2 \cite{8461368}, except that the decoder predicts discrete tokens instead of Mel-spectrograms. The rightmost decoder consists of an upsampler and a linear Transformer \cite{katharopoulos20}. Finally, the decoded tokens are rearranged and combined with the top-level tokens to be taken as input for the VQ-VAE decoder shown in Fig. \ref{fig:vqvae}.}
\label{fig:LM}
\end{figure*}

\subsection{VQ-VAE over Mel-spectrograms}
\label{ssec:vqvae}
Processing high sampling rate audio, e.g. 44.1 kHz, directly operating in the time domain is computationally demanding and often infeasible on a single GPU. Therefore, to speed up both training and inference time, our model works on Mel-spectrograms, using an additional, separate neural vocoder \cite{NEURIPS2019_6804c9bc} to convert generated mel-spectrograms back to the time-domain waveforms. Concurred with Jukebox, we find that better reconstruction quality can be obtained using hierarchical VQ-VAE \cite {razavi2019generating}. Possibly because of the difference in data representation (waveforms vs Mel-spectrograms), we do not observe codebook collapse as reported in \cite{dhariwal2020jukebox} in our work. Accordingly, we employ neither random start nor separated autoencoders.

A schematic plot of the proposed VQ-VAE architecture is shown in Fig.~\ref{fig:vqvae}. The input feature is first encoded into three levels of latent vectors by the encoder, namely bottom level, middle level, and top level. To extract coarse and fine information from the input mel-spectrogram, we use different time resolutions at different levels. The time resolution at the bottom level is the same as that of the input Mel-spectrograms, and the resolution gets halved every next level. At each level, a sequence of latent vectors $\{\mathbf{h}_l\in{R^D},l=1,...,L\}$ is extracted from the encoder head, where $D$ is the dimensionality and $L$ denotes the sequence length. A codebook that contains a set of prototype vectors $\{\mathbf{e}_j\in{R^D} ,j=1,...,M\}$ is maintained, where $M$ is the size of the book. The latent vectors $\mathbf{h}_l$ will be mapped to the nearest prototype vector in the codebook. That is, $\mathbf{h'}_l = \mathbf{e}_z$, where $ z=\arg\min_j\| \mathbf{h}_l - \mathbf{e}_j \|^2_2$. The decoder then tries to reconstruct the input Mel-spectrograms from the sequence of quantized vectors.
The VQ-VAE is trained using the following objective:
\begin{align}
    \mathcal{L} = \mathcal{L}_\text{mse} + \lambda \mathcal{L}_\text{commit} \,,
  \label{loss_total}
\end{align}
where $\mathcal{L}_\text{mse}$ denotes the \emph{reconstruction} loss of the Mel-spectrogram at the output of the decoder, $\mathcal{L}_\text{commit} = \| \mathbf{h}_l - sg(\mathbf{e}_z)\|^2_2$ is the \emph{commitment} loss encouraging the encoder to produce vectors lying close to the prototypes in the codebook, $\lambda$ is a hyperparameter that weighs the importance of the two terms, and $sg(.)$ denotes the stop-gradient operation. Noted that there are three separated commitment losses, one for the codebook in each level. During the training phase, the prototypes in the codebook are updated as a function of exponential moving averages of $\mathbf{h}$ for faster training, as suggested in the original VQ-VAE paper \cite{oord2017neural}.

As shown in Fig.~\ref{fig:vqvae}, our encoder consists of five residual blocks, each has a stack of two 1-D convolutions and a residual connection with one convolution layer. For the decoder, we use the ``G3 block'' proposed in our previous work on unconditional audio generation \cite{liu2019score} as our building component. It contains a stack of GRU, dilated convolution with feature grouping, and group normalization. Finally, the outputs from each level are followed by a convolution to project the dimensionality back to the frequency bins of the Mel-spectrogram, and are added together to reconstruct the original input. 

\subsection{Connecting Learned Tokens with Phonemes}
\label{ssec:ctc}

VQ-VAEs have proven to be able to extract high-level audio words that are closely related to phonemes, and have been used in unsupervised automatic speech recognition \cite{eloffNNGNPBWSK19,tjandra21interspeech}. Still, singing voice is vastly different from speech regarding the prosody, rhythm, and timbre. The most dominant information lying in the discrete space might not be the content of singing, and this can harm the training for the Transformer LM. Hence, we propose to minimize the CTC loss to encourage the encoded tokens to connect with the phonemes more tightly. The loss is applied to the top level encoder only, so that the content and the other information can be disentangled. That is, the top level controls the content of the singing, while the middle and bottom levels controls the style. The CTC loss is defined as the negative log likelihood of the ground truth phoneme sequence, $\mathcal{L}_\text{CTC} = -\ln p(y | \mathbf{h}_\text{top})$, where $\mathbf{h}_\text{top}$ represents the output of the encoder head at the top level, and $y$ is the ground truth phoneme sequence. Adding one hyperparameter $\alpha$, the objective then becomes:
\begin{align}
    \mathcal{L} = \mathcal{L}_\text{mse} + \lambda \mathcal{L}_\text{commit} + \alpha \mathcal{L}_\text{CTC} \,.
    \label{loss_total_ctc}
\end{align}

The original Jukebox model is trained using only the two loss terms described in Eq. \ref{loss_total}. Likely due to the difference in data representation or the size of training data (we use only 550 songs, while Jukebox uses 1.2 million songs), we found in our pilot study that these two loss terms are not sufficient to train an SVS using Mel-spectrograms. Adding the CTC loss dramatically improves the result, as we will empirically validate in the experiment section.

\subsection{Language Model: GRU \& Transformer}
\label{ssec:prior}

Once the VQ-VAE is trained, the singing voices can  be represented as three sets of discrete priors. An intuitive approach to model these distributions is via multi-stage autoregressive models as done in \cite{dhariwal2020jukebox}, starting from top to bottom one layer at a time, with the code from the preceding layer as the conditioning signal. We propose below an alternative we found more effective in our study (cf. the result of the `3-level' vs `proposed' models in Table \ref{tab:user_study}).

For modeling the top-level codes, we employ a seq2seq model to account for the input lyrics. As shown in the left and middle parts of Fig. \ref{fig:LM}, the encoder-decoder is trained to map a sequence of phonemes to a sequence of top-level codes ($\texttt{T}_1, \texttt{T}_2, ..., \texttt{T}_L$); $L$ is the length. The \emph{location-sensitive attention} \cite{chorowski2015attention}, which has been found useful for TTS \cite{8461368} (but not used in Jukebox), is used to encourage a monotonic alignment between the two sequences. Considering the context in the top level is four times smaller in size than the bottom level, we replace the top-level Transformer used in Jukebox with GRUs here for being light-weight. In our pilot study, we also found this modification is also more robust and easier to optimize.

Instead of using separate Transformers to model the middle-level codes  ($\texttt{M}_1, \texttt{M}_2, ..., \texttt{M}_\textit{2L}$) and bottom-level codes  ($\texttt{B}_1, \texttt{B}_2, ..., \texttt{B}_\textit{4L}$) as done in Jukebox, we propose to combine the codes of these two levels into a single sequence by     ``interleaving'' them and use a single Transformer to jointly model them, as shown in the rightmost part of Fig. \ref{fig:LM}.
In this way, the generation of the middle-level codes is conditioned on the whole sequence of the top-level codes, and the previously generated middle- and bottom-level codes. Namely,
\begin{align}
    p(z^\text{mid})=\prod_i p(z^\text{mid}_i \vert  z^\text{top}, z^\text{mid}_{<i}, z^\text{bot}_{<j})\,,
\end{align}
where $z^\text{top}$, $z^\text{mid}$, and $z^\text{bot}$ are the discrete codes from the top, middle, and bottom levels, respectively, and $i,j$ is the current step index of the given sequence.  We propose this setting because the output Mel-spectrogram is connected to each level of the VQ decoder via residual connections, and accordingly codes from both levels affect each other bilaterally. For implementation, we flatten the code from both levels regarding the scale factor to the top level, so the two sequences becomes one, i.e., ($\texttt{M}_1, \texttt{M}_2, \texttt{B}_1, \texttt{B}_2, \texttt{B}_3, \texttt{B}_4, \texttt{M}_3, \texttt{M}_4, \texttt{B}_5, ...$). We adopt the Linear Transformer \cite{katharopoulos20} to model the mixed sequence of middle- and bottom-level codes, for its effectiveness against autoregressive prediction on long sequences. Moreover, in light of the importance of positional embeddings \cite{vaswani2017attention} in Transformers, we propose a \emph{tick embedding} to inject the index information of the mixed sequence. The tick embedding, $tick \in \{t_1, t_2, ..., t_6\}$, where $t$ is a learnable vector, is repeated and added to the input sequence by the order of $(t_1, t_2, ..., t_6, t_1, ...)$. Finally, to condition on the top-level codes, an upsampler is developed to scale up the top-level codes to match the sequence length of mix-level prior. As observed in \cite {razavi2019generating}, using large conditioning stacks yields good performance, so we apply multiple residual block consisting of dilated convolutions in the upsampler to enlarge the receptive field.

\vspace{-2mm}
\section{Experiments}
\label{sec:experiments}
To evaluate the performance of the proposed KaraSinger model, we purchase 550 songs from an English karaoke website to build a multi-singer, accompaniment-free training dataset. 
We manually select songs from the website to include female singers and pop genre only. As all the vocals were covered by unknown amateur singers, we do not know the actual singer identity of these songs. Moreover, we observe that some of the vocals have been post-processed with reverberation and harmony, and therefore expect that the samples generated by our model may have such effects as well. We scrape the corresponding lyrics from the same website and run NUS AutoLyricsAlign \cite{9054567} on the vocals to obtain sentence-level alignments of the lyrics. (We found the alignment result reasonably accurate). We split segments that contain vocals into pieces between 5 to 15 seconds, discarding the silence. The final dataset consists of 10,589 short segments, amounting to roughly 20 hours worth of data. We reserve 100 of them as the validation set. All the singing recordings are sampled at 44.1kHz for higher quality. 

\subsection{Implementation Details}
\label{ssec:implementation}
Mel-spectrograms are extracted using a 2,048-point short time Fourier transform with a 512-point hop size, and the number of Mel bins are set to 80. We choose MelGAN \cite{NEURIPS2019_6804c9bc} for the vocoder as it has shown robustness in our preliminary experiments. Most of the settings are kept as default except a bidirectional GRU layer is added to the generator of MelGAN, and the multiplication number is doubled as the original model was used for 22.05kHz audio. It is trained for 4,000 epochs from scratch with our karaoke dataset. 

For the VQ-VAE part of our model, we found in a preliminary experiment that setting the codebook size to 256 for each of the three levels is enough to obtain satisfactory reconstruction quality. We train it for 250k iterations with batch size of 32, $\lambda$ and $\alpha$ being 0.25 and 0.10, respectively. After that, the weights of VQ-VAE are fixed and the discrete codes are extracted to prepare for training the LM. For the lyrics encoder, we use CMUDict \cite{cmudict} to convert words to their phoneme representations. During LM training, cross-entropy loss is used for optimization, and we manually inspect the generated samples to see if the quality has stopped improving (i.e., with informal listening). At last, 250k iterations were ran with batch size of 6. The Adam optimizer is used with learning rate $lr=0.0001$ for both the VQ-VAE and the LM. Nucleus sampling \cite{holtzman2019curious} is used to choose the output tokens, and the cumulative probability is set to 0.9. The total number of trainable parameters for our VQ-VAE and LM are 36M and 176M, respectively. Other details regarding the network architecture will be released in the future.

\subsection{Subjective Evaluation}
\label{ssec:subjective}
To verify the effectiveness of the proposed modifications, we conduct a listening test to evaluate the \textbf{proposed} KaraSinger model and two of its ablated versions.
\textbf{noCTC} is the case without the CTC loss during the VQ-VAE training. 
\textbf{3-level} is the case where the CTC loss is used, but the LM is built following the structure of Jukebox. That is, instead of using one Transformer decoder to jointly predict middle-level and bottom-level codes, two separate decoders are employed. After the top-level codes are generated, a 2x upsampler with a Transformer decoder is used to sample the middle-level codes, which is followed by another 2x upsampler with a Transformer to obtain the bottom-level codes. 

20 unseen sentences are selected to form the test set. Each participant is asked to rate the utterances generated by the aforementioned 3 models for in total 5 sentences from the test set, with the ordering of the 3 models randomized each time, having in mind that the samples were generated by AI models that sing the given sentence (which is shown to them) without pre-specified melody. Listening with headphone is required. The following three aspects are considered, all on a 5-point Likert scale (1--5; the higher the better):

\begin{itemize}[leftmargin=*]
    \item \textbf{Intelligibility}: 
    how well the given lyrics can be heard in the audio (i.e., how well the model is controlled by the given lyrics).
    
    \item \textbf{Musicality}: focusing on the prosody and the musical aspects of the singing, rate the quality of the sung melody.

    \item \textbf{Overall}: overall assessment considering the sound quality, pitch accuracy, pronunciation, and musicality.
    
\end{itemize}

The results from 28 anonymous participants are shown in Table \ref{tab:user_study}. It can be seen that the proposed model outperforms the other two in all the three metrics, and notably dominating the intelligibility. The model without CTC loss performs poorly, for it does not learn to attend to the input lyrics.
Our conjecture is that, given the amount of training data is extremely low compared to Jukebox, and no phoneme duration is supplied during training, it is hard to construct meaningful attention alignment by the model itself. The result of the 3-level model validates the effectiveness of predicting the audio words in the middle and bottom layers alternatively.

\begin{table}[t]
\begin{center}
\begin{tabular}{l|ccc}
\toprule
\multicolumn{1}{l|}{Model} & 
 \multicolumn{1}{c}{\bf Intelligibility} & \multicolumn{1}{c}{\bf Musicality}  &
 \multicolumn{1}{c}{\bf Overall} \\
\midrule
\textbf{noCTC}      &  1.30 $\pm$ 0.14 &    2.00 $\pm$ 0.16 &   1.64 $\pm$ 0.14 \\
\textbf{3-level}    &  3.30 $\pm$ 0.18 &    3.43 $\pm$ 0.16 &   3.19 $\pm$ 0.17 \\
\textbf{Proposed}   &  4.23 $\pm$ 0.14 &    3.85 $\pm$ 0.14 &   3.67 $\pm$ 0.15 \\
\bottomrule
\end{tabular}
\end{center}
\caption{Mean scores (plus the 95\% confidence interval) from the participants of the subjective evaluation in three different aspects.}
\label{tab:user_study}
\end{table}

\begin{figure}[t]
\includegraphics[width=0.75\linewidth,keepaspectratio]{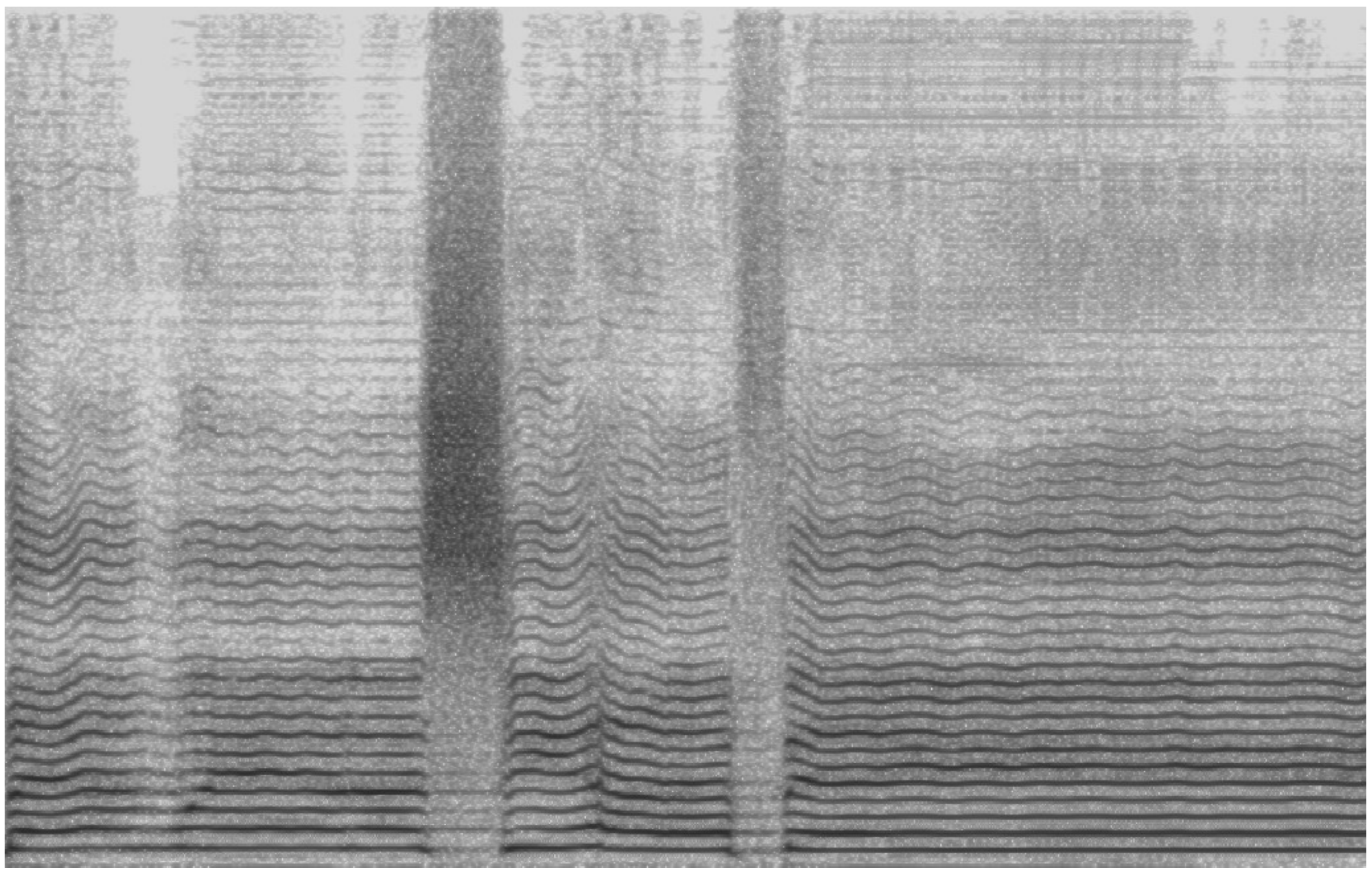}
\centering
\caption{The spectrogram of a sample generated by KaraSinger.}
\vspace{-2mm}
\label{fig:vibrato}
\end{figure}

\subsection{Qualitative Analysis}
\label{ssec:analysis}
We generate multiple samples conditioned on the same lyric that is unseen during training, finding that the result is diverse and the samples sound very differently. Moreover, we do not notice any generated melody lies similar to existing songs. 
As exemplified in Fig. \ref{fig:vibrato}, we also find that without providing any pitch-related information, our model learns to generate vibrato-like fluctuations on its own. However, the pitches are often unstable and fluctuate too much, making the generated audio sounds ``wobble'' and out of tune at times. The development of solution is left as a future work. We also generate long samples using the windowed sampling method described in the Jukebox paper \cite{dhariwal2020jukebox}. Examples of audio samples can be found on at the following demo page: \url{https://jerrygood0703.github.io/KaraSinger}.

\section{Conclusion}
\label{sec:conclusion}
We have introduced in this paper KaraSinger, an SVS model based on a hierarchical VQ-VAE over Mel-spectrograms and a lyrics-conditioned LM to achieve score-free SVS. The VQ-VAE converts existing singing recordings into sequences of discrete codes for training the LM, which can then generate novel singing audio given lyrics but no score input. 
Our subjective test shows that KaraSinger produces high-quality and natural singing voices, picking up singing techniques from the training data on its own. This opens up a new direction for SVS, where music creators can get inspirations from the synthesized singing, and common users can experience the creativity of AI. In the future, we plan to further improve the quality of the generated melody and rhythm, the long-term structure of the generated music, and explore means to further condition KaraSinger on pre-given instrumental accompaniments.

\vfill\pagebreak

\bibliographystyle{IEEEbib}
\bibliography{refs}

\end{document}